\documentclass[prl,aps,twocolumn,showpacs,superscriptaddress]{revtex4-1}
\usepackage{graphics}
\usepackage{epsfig}
\usepackage{amsmath}
\usepackage{amsfonts}
\usepackage{amssymb}
\usepackage{graphicx}
\usepackage{color}
\usepackage{tabularx}
\usepackage{txfonts}
\usepackage{soul}

\begin{document}
\title{Magnetic Order and Lattice Instabilities in  Ni$_{2}$Mn$_{1+x}$Sn$_{1-x}$ Heusler based Magnetic Shape-Memory Alloys}
\author{Vijay Singh}
\affiliation{Department of Solid-State Physics, Indian Association for the Cultivation of Science, Jadavpur, Kolkata 700032, India}
\affiliation{CEA, LITEN, 17 Rue des Martyrs, 38054 Grenoble, France}%
\author{Ambroise van Roekeghem}
\affiliation{CEA, LITEN, 17 Rue des Martyrs, 38054 Grenoble, France}%
\author{S. K. Panda}
\affiliation{Center for Advanced Materials, Indian Association for the Cultivation of Science, Jadavpur, Kolkata 700032, India}
\affiliation{Centre de Physique Th{\'e}orique, Ecole Polytechnique, CNRS UMR 7644, Universit{\'e} Paris-Saclay, 91128 Palaiseau, France}
\author{S. Majumdar}
\affiliation{Department of Solid-State Physics, Indian Association for the Cultivation of Science, Jadavpur, Kolkata 700032, India}
\author{Natalio Mingo}
\affiliation{CEA, LITEN, 17 Rue des Martyrs, 38054 Grenoble, France}%
\author{I Dasgupta}
\email{sspid@iacs.res.in}
\affiliation{Department of Solid-State Physics, Indian Association for the Cultivation of Science, Jadavpur, Kolkata 700032, India}
\affiliation{Center for Advanced Materials, Indian Association for the Cultivation of Science, Jadavpur, Kolkata 700032, India}
\date{\today}
\email{sspid@iacs.res.in}
\begin{abstract}
{The magnetic correlations in the austenite phase and the consequent martensitic transition in inverse magnetocaloric alloys, Ni$_{2}$Mn$_{1+x}$Sn$_{1-x}$, have been a matter of debate for decades. We conclusively establish using {\it ab initio} phonon calculations that the spin alignment of excess Mn at the Sn site (Mn$_{Sn}$) with the  existing Mn in the unit cell in the high temperature cubic phase of  Ni-Mn-Sn alloy is ferromagnetic (FM), and not ferrimagnetic (FI), resolving a long lasting controversy. Using first principles density functional perturbation theory (DFPT), we observe an instability of the TA$_{2}$ mode along the $\Gamma$-M direction in the FM phase, very similar to that observed in the prototypical ferromagnetic shape memory alloy (FSMA) Ni$_{2}$MnGa. This specific instability is not observed in the FI phase. Further finite temperature first principles lattice dynamics calculations reveal that at 300 K the FM phase becomes mechanically stable, while the FI phase continue to remain unstable providing credence to the fact that the high-temperature phase has FM order. These results will be primordial to understand the magneto-structural properties of this class of compounds.} 
\end{abstract}
\maketitle

\indent Transition metal based Heusler alloys, particularly those containing Mn attracted attention due to the discovery of potential half-metallic \cite{halfmetal} character in some full Heusler (Co$_2$MnGe, Co$_2$MnSi) \cite{Picozz} and semi-Heusler (NiMnSb) \cite{Ranji} compositions. Another technological interest associated with the  Mn-based full Heusler alloys is the ferromagnetic shape memory effect, which is particularly prominent in case of Ni$_2$MnGa-derived alloys \cite{Ullakko,Enkovaara}. The key to the magnetic shape memory effect is a low temperature structural transformation from a highly symmetric cubic austenitic phase to a low symmetry martensitic phase. Notably, the system remains magnetic in both phases. On top of the magnetic shape memory effect, these Heusler alloys also show an inverse magnetocaloric effect (MCE), which has its origin in the martensitic phase transformation \cite{Krenke}. In recent times there has been a considerable effort to exploit these novel structural and magnetic properties of full Heusler alloys in device applications.\\
\indent Among ferromagnetic shape memory alloys (FSMAs), Ni$_{2}$MnGa is the most intensively studied system \cite{Ullakko}. However, a major drawback of Ni$_{2}$MnGa is its brittleness. Thus, the present challenge in FSMA research lies in the search for new materials that have magneto-mechanical properties superior to Ni$_{2+x}$Mn$_{1-x}$Ga, and preferably having high martensitic start temperature (T$_{M}$) and Curie temperature (T$_{C}$). To find an alternative to Ni$_{2}$MnGa, several non-stoichiometric compositions of full-Heusler systems, {\it eg.} Ni$_{2}$MnIn and Ni$_{2}$MnSn have been studied by several groups. In this context, systems with excess Mn doping at the expense of either In ({\it i.e} Ni$_{2}$Mn$_{1+x}$In$_{1-x}$) or Sn ( {\it i.e.} Ni$_{2}$Mn$_{1+x}$Sn$_{1-x}$) are found to exhibit  martensitic phase transition (MPT) \cite{Krenke, KrenkePRB}.  Of particular interest are  Ni$_2$Mn$_{1+x}$Sn$_{1-x}$ $(0.36 \leq x \leq 0.46)$ alloys where MPT is observed in the temperature range of 100-230 K \cite{Krenke}.  It has also been reported that the crystal structures of the austenite and the martensite phases for Ni$_{2}$Mn$_{1.44}$Sn$_{0.56}$ are a Heusler-type cubic structure and an orthorhombic four layered (4O) structure, respectively \cite{Koyama1}. The magnetization in the martensite phase of this system is significantly smaller than in the austenite one. Some suitable compositions of Ni-Mn-Sn alloys are found to be promising inverse magneto-caloric materials where entropy increases upon application of magnetic field. In order to disentangle the magnetic contribution to the entropy change an accurate knowledge of magnetic ordering during structural transition is necessary. There  have been vigorous experimental and theoretical activities to understand the magnetic ordering, in particular the magnetic correlation between Mn and  Mn$_{Sn}$ and also to understand the underlying mechanism of the MPT in this class of non-stoichiometric systems. \\

\indent The nature of magnetic interaction between Mn and Mn$_{Sn}$ (FM vs FI), has been a subject of controversy for a long time. Neutron Diffraction \cite{Brown1}, Neutron Polarization analysis and magnetization studies \cite{Aksoy1} emphasize that in the austenite phase the magnetic correlations between Mn and Mn$_{Sn}$ are ferromagnetic. However several magnetization studies by other groups( \cite{Cakir} and references therein)  suggest that the magnetic correlations between the Mn and Mn$_{Sn}$ are FI . On the other hand, first principles electronic structure calculations based on density functional theory (DFT) by several groups \cite{Siewert2, Soko, Ye, Li} argue that this magnetic correlation is predominantly FI. It is important to note that the austenite phase is a high temperature phase, and DFT is well known to be particularly suitable for ground state properties of materials. Thus, the reliability of the DFT calculations for the high temperature austenite phase remains questionable. 

\indent In addition to the nature of magnetic order, the origin of the observed martensitic transformation in Ni-Mn-Sn alloys has also been under intense debate for a long time. For example, using hard X-ray photoelectron spectroscopy measurements, Ye {\it et al.} have suggested that Jahn-Teller splitting of the Ni-3d e$_{g}$ state plays an important role in driving the instability of the cubic phase for $x$ $\leq$ 0.36 \cite{Ye}. However, in contrast to these results, a very recent work calculated the relative on-site energies of Ni d states using tight binding model for the Ni$_{2}$Mn$_{1.5}$Sn$_{0.5}$ alloy and found that these energies are small. Based on these results, the authors ruled out the possibility of Jahn-Teller distortions {\cite{Pal} and  suggested that Ni-Mn hybridization and impact of Sn lone pair is responsible for the MPT. Furthermore, extended X-ray absorption fine-structure (EXAFS) measurements suggest that local structural distortions upon excess Mn doping play a crucial role for the austenite to martensitic phase transition \cite{Bhobe1}. Using both time of flight neutron spectroscopy and {\it ab-initio} calculations Recarte {\it et al.} suggest a predominant role of the vibrational entropy in driving the martensitic transformation \cite{Recarte}. There are however no systematic phonon calculation available for Ni$_{2}$Mn$_{1+x}$Sn$_{1-x}$ systems. Thus, the origin of the structural transformation also remains an unsolved problem in non-stoichiometric Ni-Mn-Sn alloys.\\ 
In this paper, we shall present electronic, magnetic, elastic and vibrational properties for the parent compound Ni$_{2}$MnSn and for Ni$_{2}$Mn$_{1+x}$Sn$_{1-x}$ ($x$ = 0.5) in order to clarify the nature of the magnetic interaction between Mn and Mn$_{Sn}$ and to understand the origin of the martensitic transition in such non-stoichiometric compositions.

\indent The crystal structure of the stoichiometric Ni$_{2}$MnSn in the high temperature L$_{2_{1}}$ cubic full Heusler phase is shown in SM \cite{SM}, Fig. S1(a). The system is ferromagnetic with Curie temperature T$_{c}$, 362 K \cite{Planes1}. The non-stoichiometric systems with excess Mn at the Sn site Ni$_{2}$Mn$_{1.48}$Sn$_{0.52}$  also form a cubic structure in the high temperature austenite phase \cite{Aksoy1,Acet1,Acet2}. In order to simulate the excess Mn at the Sn site, close to the experimental stoichiometry, we created  a supercell that is four times the original unit cell, and replaced some of the Sn atoms by Mn atoms (Mn$_{Sn}$). We have considered two kinds of magnetic configurations, namely the FM and FI phase as shown in the Fig. S1(b) and (c) of SM \cite{SM}, respectively. 

\indent All the electronic structure calculations are performed in a plane-wave basis set using the projector augmented wave (PAW) method as implemented in the Vienna {\it ab-initio} simulation package (VASP) \cite{ Blochl, Kresse1, Kresse2}. The plane-wave cutoff energy was chosen to be 500 eV and a 4$\times$4$\times$4 Monkhorst-Pack grid was employed to sample the Brillouin zone (BZ).  The exchange-correlation part was approximated by the PBE functional within the generalized gradient approximation (GGA) \cite{PBE}. \\
\indent We have studied the lattice-dynamical properties using the density functional perturbation theory (DFPT) as implemented in  the  Plane-Wave Self-Consistent Field (PWSCF) code. This code allows to calculate the dynamical matrix at any q point in the BZ directly \cite{Giannozzi}. The energy threshold value for the convergence is taken to be 10$^{-14}$ Ryd in phonon calculations. Dynamical matrices are calculated with a k-point grid of 4$\times$4$\times$4 and 2$\times$2$\times$2 in the irreducible Brillouin zone for the one formula unit FCC and four formula unit simple cubic structures, respectively (see SM \cite{SM} for details). 
\indent We have also performed finite temperature lattice dynamics calculations using our recently developed inhouse code\cite{ScF2}. In this method the temperature dependent interatomic force constants are calculated employing a regression analysis of forces obtained from DFT calculation coupled with a harmonic model of the quantum canonical ensemble. The calculations are done in an iterative way to achieve self-consistency of the phonon spectrum (see SM \cite{SM}).

\indent To investigate the role of magnetic order (FM and FI) in triggering the martensitic transformation in the Ni-Mn-Sn alloy we carried out first principles DFPT calculations to investigate possible phonon anomalies in the FM state for the stoichiometric Ni$_{2}$MnSn and both in the FM and FI configuration in the  non-stoichiometric Ni$_{2}$Mn$_{1+x}$Sn$_{1-x}$, $x$ = 0.5 alloy.  For the purpose of comparison, we first show the phonon dispersion for Ni$_{2}$MnGa ($a$ = 5.825 $\AA$) and Ni$_{2}$MnSn ($a$ = 6.05 $\AA$) along $\Gamma$-X for one formula unit FCC unit cell, (see SM, Fig. S2 (a) \cite{SM}).  The phonon dispersion for Ni$_{2}$MnGa shows a dip in the lowest branch of the transverse acoustic mode (TA$_{2}$). Thus, the Ni$_{2}$MnGa crystal is dynamically unstable to the lattice distortion corresponding to the eigenvector of this mode. This result is in agreement with earlier calculations on Ni$_{2}$MnGa \cite{Zayak}. The origin of the anomaly of the phonon dispersion was attributed to the Kohn anomaly due to Fermi surface nesting and electron-phonon interaction \cite{Rabe}. Interestingly, such a softening of the phonon mode is not seen for Ni$_{2}$MnSn, consistent with the fact that stoichiometric Ni$_{2}$MnSn does not exhibit a martensitic transition (see SM, Fig. S2 (b) \cite{SM}).\\

\begin{figure}[t]
\includegraphics[width=0.99\columnwidth]{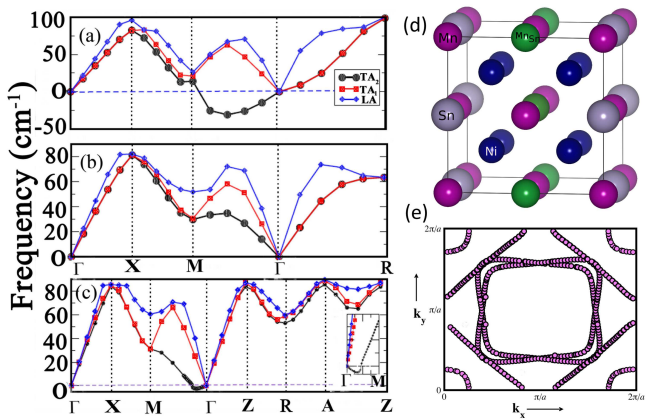}
\caption{Phonon dispersion (only acoustic modes) of (a) ferromagnetic Ni$_{2}$MnGa (b) ferromagnetic Ni$_{2}$MnSn and (c) FM state Ni$_{2}$Mn$_{1.5}$Sn$_{0.5}$ along high-symmetry lines of BZ (inset dispersion along $\Gamma$-M). The wave vector coordinate is in units of ($\frac{2\pi}{a}$).(d) The four formula unit supercell employed for phonon calculations (e)Cross-section of the minority-spin Fermi surface (k$_{z}$ = 0) for FM state of Ni$_{2}$Mn$_{1.5}$Sn$_{0.5}$. Nesting is seen along vector (1,1,0)2$\pi$/a direction. }
\label{1}
\end{figure}

\indent Next, we have considered Ni$_{2}$Mn$_{1.5}$Sn$_{0.5}$ in the FM phase. As discussed earlier, this structure was simulated using a four formula unit supercell. Substitution of Mn at the Sn site breaks the local cubic symmetry to a tetragonal one. Firstly, for the purpose of comparison we have shown the phonon dispersion for Ni$_{2}$MnGa and Ni$_{2}$MnSn in the ferromagnetic structure, but have now considered a four formula unit supercell (See Fig. 1(d)). As expected there are imaginary frequencies associated with the TA$_{2}$ mode for Ni$_{2}$MnGa and no such imaginary frequencies are seen for Ni$_{2}$MnSn (Fig. 1a-b). For Ni$_{2}$MnGa the observed imaginary frequencies are now along $\Gamma$-M due to the folding of the phonon dispersion in the smaller BZ of the four formula unit supercell. This clearly demonstrate the accuracy and the robustness of our results even for the bigger supercell. 

\indent Next, we have calculated the phonon dispersion for Ni$_{2}$Mn$_{1.5}$Sn$_{0.5}$. Interestingly, Ni$_{2}$Mn$_{1.5}$Sn$_{0.5}$ shows a phonon softening of  the TA$_{2}$ mode similar to Ni$_{2}$MnGa, Fig. 1c. The full phonon dispersion for the FM phase Ni$_{2}$Mn$_{1.5}$Sn$_{0.5}$ is presented in SM (Fig. S3(a)) \cite{SM}. A comparison of Fig 1(a) and Fig 1(c) confirm that  Ni$_{2}$Mn$_{1.5}$Sn$_{0.5}$ crystal is dynamically unstable as in Ni$_{2}$MnGa \cite{Rabe} . This result indicates that Ni$_{2}$Mn$_{1.5}$Sn$_{0.5}$ not only exhibit a martensitic transition, but also the mechanism for the MPT is very similar to that for Ni$_{2}$MnGa \cite{Manosa}.\\

\indent Similarly as in Ni$_{2}$MnGa, it is expected that a nesting feature \cite{Rabe} is also responsible for the imaginary frequencies for the phonon mode in the ferromagnetic Ni$_{2}$Mn$_{1.5}$Sn$_{0.5}$.} In order to examine this scenario we plotted the Fermi surface in the minority spin channel in the k$_{z}$ = 0 plane (Fig. 1(e)). We observe a large as well as a small region of the Fermi surface exhibiting  nesting along q=$\frac{2\pi}{a}$(1,1,0) direction of the BZ. This causes a reduction of phonon energies in a narrow range of wavevector, which is indeed necessary for the structural transition. 

\indent  The calculated phonon dispersion relations also allow one to estimate the elastic constants based on sound velocity propagating in the crystal. The results of our calculation are summarized in Table 1.It is well known that martensitic transformations (MTs) are accompanied by elastic modulus softening \cite{Nakanishi}. As expected, we also find  the softening of c$^{\prime}$ = 1.33 GPa ({\it i.e.} modulus of the tetragonal shear distortion). It reduces drastically for the non-stoichiometric Ni-Mn-Sn alloy in the FM phase in comparison to the parent compound Ni$_{2}$MnSn ($x$ = 0) where c$^{\prime}$ = 32.15 GPa. A comparison with Ni$_{2}$MnGa shows excellent agreement between the two systems. The near zero value of c$^{\prime}$ implies that there is almost no energy cost for a small tetragonal shear distortion of the type $2\epsilon_{\alpha\alpha}=2\epsilon_{yy}=-\epsilon_{zz}$. \\  
\indent On the other hand, the phonon spectra for the FI structure does not lead to softening or imaginary frequencies for the acoustic mode, but there are imaginary frequencies associated with the optical mode implying that the FI phase is unstable, but probably does not lead to the martensitic transition(See SM, Fig. S3(b) in \cite{SM}). As expected, a typical Fermi surface nesting feature is also absent in the FI phase (SM, See Fig. S4(e,f))\cite{SM}. All these results point to a single conclusion that martensitic transition in Ni$_{2}$Mn$_{1.5}$Sn$_{0.5}$ is only possible provided the magnetic interactions are ferromagnetic.\\

\bgroup
\def\arraystretch{1.2}
\begin{table}[t]
\caption{Elastic constants, in GPa, for ferromagnetic Ni$_{2}$MnSn as well as  FM phase of Ni$_2$Mn$_{1.5}$Sn$_{0.5}$ alloy. For Ni$_{2}$MnGa, results from previous theoretical study~\cite{Zayak1} and experiment~\cite{Worgull} are shown for comparison.}
\centering \vspace{0.1cm}
\begin{tabular}{cccccc}
\hline
\hline
\hspace{0.15cm} Compounds \hspace{0.15cm} & \hspace{0.15cm} C$_{11}$ \hspace{0.15cm} & \hspace{0.15cm} C$_{12}$ \hspace{0.15cm} & \hspace{0.15cm} C$_{44}$ \hspace{0.15cm} & \hspace{0.15cm} C$^{\prime}$ \hspace{0.15cm} & \hspace{0.15cm} C$_{L}$ \hspace{0.15cm} \\
\hline 
Ni$_{2}$MnGa (this work) & 164.0 & 134.2 & 90.8 & 14.9 & 239.9 \\
Ni$_{2}$MnGa (Ref.~\onlinecite{Zayak1}) & 139.4 & 122.6 & 91.0 & 8.2 & 222.0 \\
Ni$_{2}$MnGa (Ref.~\onlinecite{Worgull}) & 152.0 & 143.0 & 103.0 & 4.5 &  250.0 \\
Ni$_{2}$MnSn & 166.3 & 101.9 & 82.2 & 32.2 & 216.3 \\
Ni$_{2}$Mn$_{1.5}$Sn$_{0.5}$ & 129.7 & 127.0 & 81.3 & 1.3 & 209.7 \\
\hline
\hline
\end{tabular}
\label{6.III}
\end{table}
\egroup


\indent To substantiate, we have carried out finite temperature self-consistent lattice dynamical calculations on non-stoichiometric Ni$_{2}$Mn$_{1+x}$Sn$_{1-x}$ $x$=0.5, for both FM and FI phases in the cubic structure, (see Fig. 2 and Fig. S5 \cite{SM}). Our calculations reveal that the high temperature structure is mechanically stable provided the magnetic correlations between Mn and Mn$_{Sn}$ are FM, while the FI state is masked with imaginary frequencies making the structure unstable. This is in sharp contrast with the available DFT calculations \cite{Ye, Siewert2} which invariably find FI state to be energetically favorable. This suggests that  DFT calculations may not be adequate to understand the magnetic properties of the high temperature austenite phase.  The full phonon dispersion for both FM and FI phase, obtained at 300 K, is shown in SM (Fig. S5 \cite{SM}).

\indent Having established that the ferromagnetic alignment of spins in the austenite phase is responsible for the martensitic transition we have next investigated the alignment of excess Mn spins at the Sn site in the martensitic phase.  We have calculated the total energy as a function of c/a using GGA functional where the volume of the cell is kept fixed. The variation of the total energy with respect to the tetragonal distortion from the cubic phase is calculated for Ni$_{2}$Mn$_{1.5}$Sn$_{0.5}$ and the results are shown in Fig. 3(a). Our results are in excellent agreement with previous studies Ref.\cite{Ye, Wang1, Wang2}. For c/a $>$ 1, the FI state has the lowest energy, indicating not only martensitic instability, but also providing clear evidence that excess Mn ions at the Sn sites (Mn$_{Sn}$) are antiparallel to the already existing Mn ions in the unit cell in the martensitic phase. This explains the reduction of the magnetic moment in the martensitic phase in comparison to the austenite phase observed experimentally. Moreover local minima seen for c/a $<$ 1 for the Ni$_{2}$Mn$_{1.5}$Sn$_{0.5}$ system in both the FM and FI states (see Fig. 3a ) suggests the development of a modulated structure as discussed by Siewart {\it et al.} \cite{Siewert} , Himmetoglu {\it et al.} \cite{Him} and Ayuela {\it et al.} \cite{Ayuela} in the context of Ni$_{2}$MnGa.\\

\begin{figure}[t]
\includegraphics[width=0.99\columnwidth]{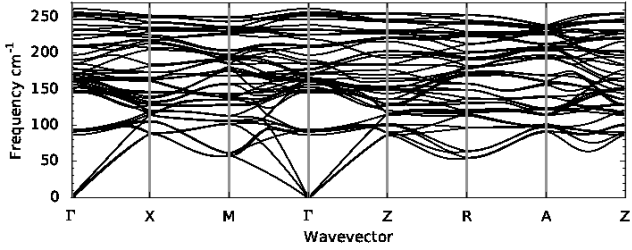}
\caption{Full phonon dispersion of Ni$_{2}$Mn$_{1.5}$Sn$_{0.5}$ along the high-symmetry lines of the BZ for the  ferromagnetic state at 300 K. The wave vector coordinate is in units of (2$\pi$/a).}
\label{2}
\end{figure}

\indent In order to identify the signatures both in the crystal structure and electronic structure in the low temperature cubic phase which will alleviate upon MPT, we first discuss the local structural  distortion in Ni-Mn-Sn alloy taking into account both the FM ad FI configurations. Table S2 of the SM \cite{SM} displays, the various bond lengths calculated for the unrelaxed and the relaxed cases (until all atomic forces are below 0.001 eV/$\AA$) using the GGA method. From Table S2 , we gather that there is local lattice distortion where the distance between the Ni-Mn, Ni-Mn$_{Sn}$, as well as Mn-Mn$_{Sn}$ are reduced {\it  irrespective of whether the correlation between Mn and Mn$_{Sn}$ is FM or FI}. Our calculated results are consistent with recent EXAFS spectroscopy. \cite{Bhobe1} This in turn enhances the hybridization between the Ni-d and Mn-d states. Further, the Mn-Mn$_{Sn}$ distance is appreciably lower in comparison to the Mn-Mn distance in the stoichiometric systems. This, in turn, will have impact on the exchange interactions. In particular short  Mn - Mn$_{Sn}$ distance will promote antiferromagnetic interactions due to the fact that Mn-Mn exchange interactions are primarily mediated by conduction electrons and exhibit long range oscillating behavior being AFM at short distances. As a consquence FI state will  stabilize in the low temperature martensitic phase \cite{Bruno}.

\begin{figure}[t]
\includegraphics[width=0.85\columnwidth]{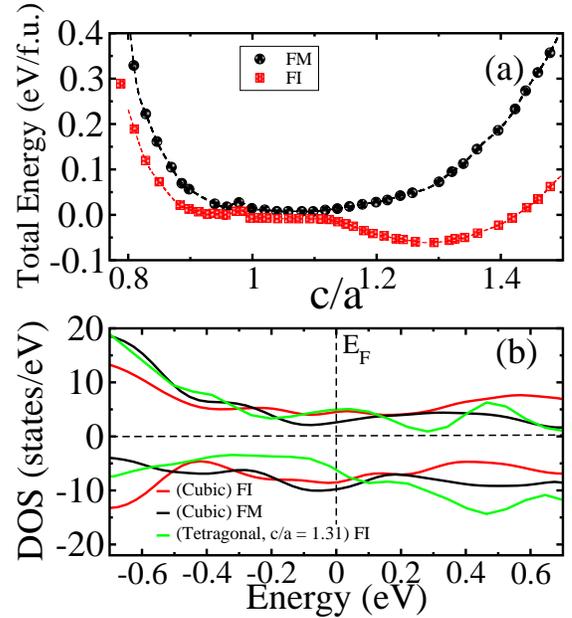}
\caption{(a) Total energy difference (per formula unit) relative to the cubic phase as a function of the axial ratio, c/a, in the tetragonal structure calculated for Ni$_{2}$Mn$_{1.5}$Sn$_{0.5}$ for parallel and anti-parallel magnetic interactions for GGA method. (b) Spin polarized total DOS in the cubic phase of Ni$_{2}$Mn$_{1.5}$Sn$_{0.5}$ for both FM and ferrimagnetic (FI) configurations. The DOS in the tetragonal phase (c/a = 1.31) for Ni$_{2}$Mn$_{1.5}$Sn$_{0.5}$ in ferrimagnetic configuration is also shown for comparison.}
\label{3}
\end{figure}

\indent Finally, to understand the electronic structure of the non-stoichiometric system we have displayed the total DOS, in the FM and FI spin configuration for the cubic austenite phase and in the FI configuration for the tetragonal phase in Fig. 3b. We find that the weight of the DOS  at the Fermi level in the minority spin channel for the FI configuration is reduced in comparison to the FM and FI configuration in the cubic phase, adding to the structural stability upon MPT.

\indent In this paper, we have investigated in detail the electronic structure, magnetism  and phononic properties of Ni$_{2}$Mn$_{1.5}$Sn$_{0.5}$ alloy, emphasizing that the nature of the magnetic configuration in the high temperature austenite phase is intimately related to the low temperature structural phase transition. 
The phonon dispersion calculated for the FM phase for Ni$_{2}$Mn$_{1.5}$Sn$_{0.5}$ presents imaginary frequencies associated with the TA$_{2}$ mode, analogous to that observed for Ni$_{2}$MnGa,which are responsible for the martensitic transition. The origin of the martensitic transition is attributed to the nesting of the Fermi surface in the minority spin channel. No such phonon anomalies were observed for the FI phase. Using finite temperature lattice dynamic calculations, we have shown that at 300 K, the imaginary frequencies associated with the TA$_{2}$ mode along the  $\Gamma$–M direction of the BZ disappear only for FM phase. This clearly rules out the possibility of the FI phase in the high temperature cubic austenite phase. 
 The variation of the total energy with respect to the distortion from the cubic phase yields a minimum at $\frac{c}{a} > 1$ only for the FI state for Ni$_{2}$Mn$_{1.5}$Sn$_{0.5}$, providing a clear evidence that in the martensite phase the correlation between Mn and Mn$_{Sn}$ are antiparallel. Our calculations provide a natural explanation of the reduction in the magnetic moment observed in the martensite phase. Finally we have discussed the signatures in the crystal structure and electronic structure in the cubic phase at low temperature which are eventually alleviated upon MPT.  \\

We wish to thank Prof. Dan Thomas Major for an insight which led to an important improvement in this work. ID thanks Technical Research Centre, Department of Science and Technology Govt. of India for support.

\end{document}